\documentclass[reprint,prb,aps,superscriptaddress,amsmath]{revtex4-1}

\usepackage{graphicx}
\usepackage{txfonts}
\usepackage{color}
\usepackage{dcolumn}
\usepackage{mathrsfs}

\begin{document}
\title{Occupied-orbital fast multipole method for efficient exact exchange evaluation} 
\author{Hai-Anh Le}
\email{anh@u.northwestern.edu}
\author{Toru Shiozaki}
\affiliation{Department of Chemistry, Northwestern University, 2145 Sheridan Rd., Evanston, IL 60208}
\date{\today}

\begin{abstract}
We present an efficient algorithm for computing the exact exchange contributions in the Hartree--Fock and hybrid density functional theory models 
on the basis of the fast multipole method (FMM).
Our algorithm is based on the observation that
FMM with hierarchical boxes can be efficiently used in the exchange matrix construction,
when at least one of the indices of the exchange matrix is constrained to be an occupied orbital.
Timing benchmarks are presented for alkane chains (C$_{400}$H$_{802}$ and C$_{150}$H$_{302}$),
a graphene sheet (C$_{150}$H$_{30}$), a water cluster [(H$_2$O)$_{100}$],
and a protein Crambin (C$_{202}$H$_{317}$O$_{64}$N$_{55}$S$_6$).
The computational cost of the far-field exchange evaluation for Crambin is roughly 3\% that of a self-consistent field iteration 
when the multipoles up to rank 2 are used.
\end{abstract}

\maketitle

\section{Introduction}
Evaluating the electron--electron interaction in mean-field models, such as the Hartree--Fock method and hybrid density functional theory,
is challenging because the bare electron--electron interaction is long range.
This is in contrast to the evaluation of screened interactions in dynamical electron correlation problems,
which has been resolved to a great extent by local correlation approaches.\cite{pulay83cpl,schutz01jcp,riplinger16jcp}
In particular, computation of the exact exchange contributions in the mean-field models remains an important challenge in quantum chemistry.\cite{rebolini16jctc}
The exchange matrix elements are defined as
\begin{align}
&K_{rs} = \sum_{tu} (rt|us) D_{tu} = 2\sum_{i}(ri|iu),\\
&(rs|tu) = \iint d\mathbf{r}_1d\mathbf{r}_2\phi_r(\mathbf{r}_1)\phi_s(\mathbf{r}_1)
           \frac{1}{r_{12}}
           \phi_t(\mathbf{r}_2)\phi_u(\mathbf{r}_2),
\label{erib}
\end{align}
where $r$, $s$, $t$, and $u$ label atomic orbitals (AOs). Hereafter $i$ and $j$ label occupied orbitals.
$D_{tu}$ are the density matrix elements, which become diagonal in the canonical molecular orbital (MO) representation.
There have been extensive studies to optimize the exchange evaluation: for instance, 
the LinK method,\cite{ochsenfeld98jcp,ochsenfeld00cpl}
multipole accelerated algorithms,\cite{burant96jcp, schwegler99jcp}
rigorous integral screening,\cite{schwegler96jcp, schwegler97jcp, maurer12jcp,maurer13jcp}
density fitting with local domains,\cite{polly04mp,koeppl16jctc}
truncated or short-range exchange kernels with and without the use of the resolution-of-the-identity (RI) approximation,
\cite{jung05pnas, izmaylov06jcp, paier09prb, guidon09jctc}
the chain-of-sphere exchange method based on quadrature,\cite{neese09cp}
the auxiliary density matrix method,\cite{guidon10jctc}
the pair-atomic RI approximation,\cite{merlot13jcc,hollman14jcp,manzer15jctc,manzer15jcp}
and the low-rank decomposition of the exchange operator.\cite{lewis16jctc, lin16jctc}

In this work, we report an efficient algorithm, termed occupied-orbital fast multipole method for exchange (occ-FMM-K), for computing the exchange contributions based on the fast multipole method (FMM).\cite{greengard87thesis, greengard87jcp, greengard94science}
Despite its tremendous success in Coulomb matrix construction,
\cite{white94jcp, white94cpl, petersen94jcp, kutteh95cpl, strain96science,perez-jorda96jcp,kudin97cpl,
challacombe97jcp2, kudin04jcp, choi01jcc, watson04jcp, rudberg06jcp, toivanen15pccp}
FMM has been considered inapplicable to efficient computation of far-field exchange interactions.
Our FMM-based algorithm for the exact exchange contributions neither relies on local orbitals nor introduces numerical truncation (see below),
making it amenable for future extensions of the algorithm to extended systems with small band gaps and to efficient computation of response properties.

FMM was first introduced a few decades ago for evaluating the far-field Coulomb interaction energies
between classical charges.\cite{greengard87thesis, greengard87jcp, greengard94science}
Many quantum chemical programs have since been developed for the Coulomb matrix evaluation.%
\cite{white94jcp, white94cpl, petersen94jcp, kutteh95cpl, strain96science,perez-jorda96jcp,kudin97cpl,
challacombe97jcp2, kudin04jcp, choi01jcc, watson04jcp, rudberg06jcp, toivanen15pccp}
In FMM, one approximates the two-electron Coulomb operator for separated charge distributions using the scaled regular and irregular solid harmonics,
\begin{align}
\frac{1}{r_{12}} &= \sum_{ll'mm'}(-1)^l O_{l,m}(\mathbf{r}_1 - \mathbf{X}) 
M_{l+l',m+m'}(\mathbf{X}-\mathbf{X}') \nonumber\\
&\quad \times O_{l',m'}(\mathbf{r}_2 - \mathbf{X}'), 
\label{mmexp}
\end{align} 
in which the factor $(-1)^l$ arises from the parity of the associated Legendre polynomials.
The scaled regular and irregular solid harmonics (often referred to as multipoles and local expansions) are defined as 
\begin{subequations}
\begin{align}
&O_{l,m}(\mathbf{r}) = \epsilon_m\frac{r^l}{(l+|m|)!}P_{l,|m|}(\cos\theta)e^{-im\phi},\\
&M_{l,m}(\mathbf{r}) = \epsilon_m\frac{(l-|m|)!}{r^{l+1}} P_{l,|m|}(\cos\theta)e^{im\phi},
\end{align}
\end{subequations}
in which $\mathbf{r}$ is written in spherical coordinates $(r, \theta, \phi)$ on the right-hand side.
$P_{l,m}$ is the associated Legendre polynomial, and $\epsilon_m$ is a phase factor that is $1$ if $m\ge 0$ and $(-1)^m$ otherwise.

The main idea of this work is to use 
FMM for computing only the exchange matrix elements that have (at least) one occupied-orbital index,
taking advantage of the fact that
the only matrix elements that are required to find the mean-field solution are $K_{ri}$, 
\begin{align}
K_{ri} = 2\sum_j (rj|ji).
\label{occex}
\end{align}
In other words, the virtual--virtual block of the exchange matrix $K_{ab}$ is not strictly necessary.
This is because self-consistent solutions minimize the mean-field energy with respect to
orbital rotations between occupied and virtual orbitals $\kappa_{ia}$, at which the following energy gradients are made zero:
\begin{align}
\frac{\partial E}{\partial \kappa_{ia}} = 2\left(h_{ia} + J_{ia} - \frac{1}{2} K_{ia}\right).
\end{align}
Note that the final energy can be computed from $K_{ij}$.
The use of the partial exchange matrix [Eq.~\eqref{occex}] has been reported in a recent work by Manzer et al., who have introduced the so-called occ-RI-K algorithm.\cite{manzer15jcp}
As shown below, this trick is essential for utilizing a hierarchy of boxes with upward and downward translation of multipoles in the FMM algorithm.
We report an efficient, parallel implementation of the algorithm, which is publicly available as part of the {\sc bagel} package.\cite{bagel,bagelreview}

\begin{figure*}[tb]
\includegraphics[width=\textwidth,keepaspectratio]{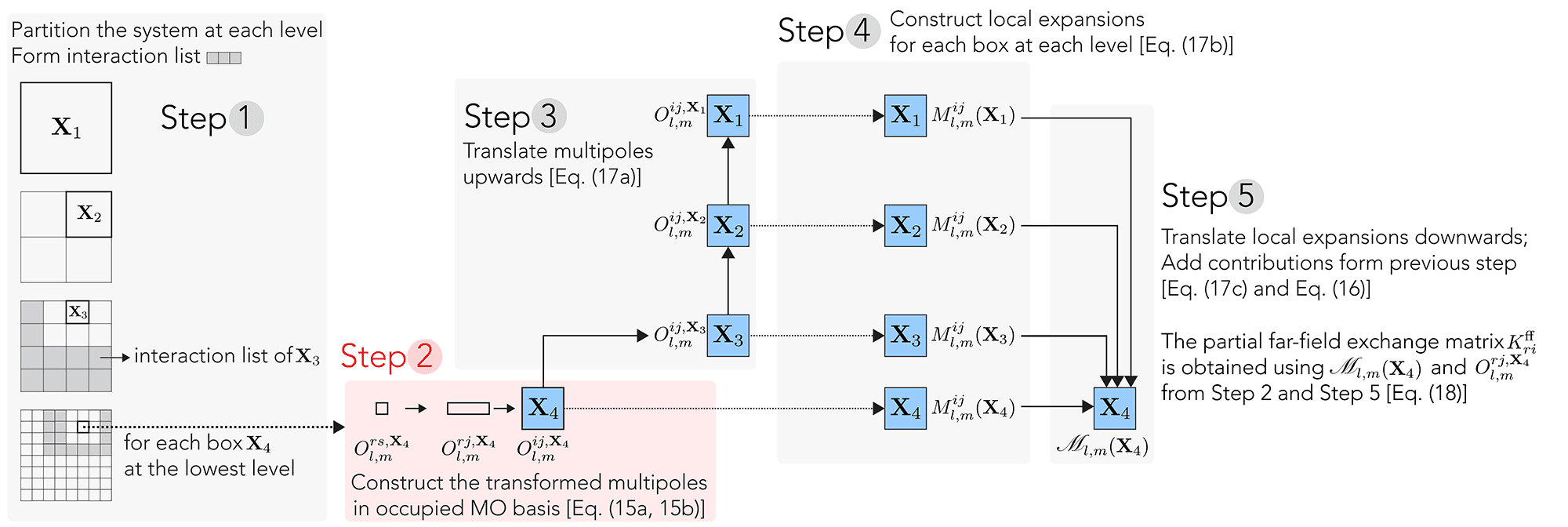}
\caption{Schematic representation of the occ-FMM-K algorithm for constructing $K_{ri}^\mathrm{ff}$
         using the translation relations up and down the FMM hierarchy. Step 2 is the essence of 
         occ-FMM-K.\label{algfig}}
\end{figure*}

\section{Theory}
When two basis-function pairs, $\phi_r(\mathbf{r}_1)\phi_s(\mathbf{r}_1)$ and $\phi_t(\mathbf{r}_2)\phi_u(\mathbf{r}_2)$, are sufficiently separated,
the electron repulsion integrals [Eq.~\eqref{erib}] can be approximated using Eq.~\eqref{mmexp} as
\begin{align}
(rs|tu) &= \sum_{lm} (-1)^l O_{l,m}^{rs,\mathbf{X}}\sum_{l'm'}M_{l+l',m+m'}(\mathbf{X}-\mathbf{X}')O_{l',m'}^{tu,\mathbf{X}'}.
\label{eri}
\end{align}
The multipole integrals over atomic-orbital basis functions are defined as
\begin{align}
O_{l,m}^{rs,\mathbf{X}} = \int d\mathbf{r} \phi_r(\mathbf{r}) {O}_{l,m}(\mathbf{r}-\mathbf{X}) \phi_s(\mathbf{r}).
\end{align}
The approximated integrals Eq.~\eqref{eri} now depend only on the multipole integrals and the separation between the expansion centers,
$\mathbf{X}$ and $\mathbf{X}'$.
In our FMM implementation, the expansion centers are taken to be the center of the Cartesian box to which the basis pair belongs (note that $\mathbf{X}$ is unique to each pair of $r$ and $s$).

When constructing the Coulomb matrix, the Coulomb potential at center $\mathbf{X}$, $\mathscr{M}_{l,m}(\mathbf{X})$, due to the charge distributions
associated with all distant basis function pairs $\phi_t(\mathbf{r}_2)\phi_u(\mathbf{r}_2)$
is evaluated as follows.
First, we contract
the density matrix elements $D_{tu}$ and
the multipoles associated with $\phi_t(\mathbf{r}_2)\phi_u(\mathbf{r}_2)$ that are centered at $\mathbf{X}'$ 
($O_{l',m'}^{tu,\mathbf{X}'}$) to define
multipole tensors $O_{l',m'}^{\mathbf{X}'}$ for each box containing these distributions. Then,
these multipole tensors are multiplied by the local expansions to give $\mathscr{M}_{l,m}(\mathbf{X})$,
\begin{align}
\mathscr{M}_{l,m}(\mathbf{X}) = (-1)^l\sum_{\mathbf{X}'}\sum_{l'm'}M_{l+l',m+m'}(\mathbf{X}-\mathbf{X}') O_{l',m'}^{\mathbf{X}'},
\label{fmmc}
\end{align}
Note that the quantity $\mathscr{M}_{l,m}(\mathbf{X})$ is defined for each of the boxes
and includes all the distant Coulomb interactions.
Using this, the far-field part of the Coulomb matrix is computed as
\begin{align}
J_{rs}^\mathrm{ff} = \sum_{lm} O_{l,m}^{rs,\mathbf{X}}\mathscr{M}_{l,m}(\mathbf{X}).
\end{align}
The Coulomb matrix elements associated with the neighboring charge distributions, {\it i.e.}, the near-field region,
where the multipole expansion is no longer valid, are evaluated using standard algorithms.
The cost of computing the near-field part is linearly scaling with respect to system size.

The summation in Eq.~\eqref{fmmc} is efficiently performed using a hierarchy of boxes that are
constructed by first partitioning the system of interest into a number of boxes, each of which is further divided
into smaller boxes and so on. This hierarchical structure allows the distant contributions to be computed
at the coarse-grained levels (or higher levels, with larger and fewer boxes),
and translated to the lower levels using the spherical harmonics addition theorem for scaled regular and irregular solid harmonics.
The standard FMM algorithm consists of three steps: First, 
the multipoles are computed at the lowest level and translated upward,\cite{greengard87thesis,white94jcp,choi01jcc} 
\begin{align}
O_{l,m}^{\mathbf{X}_p} &= \sum_{l'm'} O_{l-l',m-m'}^{\mathbf{X}_c - \mathbf{X}_p} O_{l',m'}^{\mathbf{X}_c},
\label{fmm1}
\end{align}
where $\mathbf{X}_p$ and $\mathbf{X}_c$ are the center of the parent and child boxes, respectively, and $l\le L_\mathrm{max}$.
Second, the local expansions are obtained by translating the multipoles within the same level.
To do so, each box has an interaction list that enumerates the non-neighboring boxes at the same level
whose parents are its parent's neighbor. It reads\cite{white94jcp,choi01jcc} 
\begin{align}
\mathscr{M}_{l,m}(\mathbf{X}) &= (-1)^l \sum_{l'm'}M_{l+l',m+m'}(\mathbf{X} - \mathbf{X}') O_{l',m'}^{\mathbf{X}'},
\label{fmm2}
\end{align}
in which $\mathbf{X}$ and $\mathbf{X}'$ are the center of the box and that associated with those in the interaction list, respectively.
Finally, the local expansions are translated downward,\cite{greengard87thesis,white94jcp,choi01jcc}
\begin{align}
\mathscr{M}_{l,m}(\mathbf{X}_c) &= \sum_{l'm'}M_{l',m'}(\mathbf{X}_p) O_{l'-l,m'-m}^{\mathbf{X}_c - \mathbf{X}_p}.
\label{fmm3}
\end{align}
The local expansions containing the far-field interactions for all of the boxes at the lowest level
are then collected to construct $\mathscr{M}_{l,m}(\mathbf{X})$ in Eq.~\eqref{fmmc}.

In this work, we have extended this algorithm to computation of partial exchange matrix elements [Eq.~\eqref{occex}].
Our new algorithm is termed occ-FMM-K.
The molecular integrals that contribute to the occupied exchange matrix are written using the multipole approximation as
\begin{align}
(rj|ji) &= \sum_{lm} (-1)^l O_{l,m}^{rj,\mathbf{X}}\sum_{l'm'}M_{l+l',m+m'}(\mathbf{X}-\mathbf{X}')O_{l',m'}^{ji,\mathbf{X}'}.
\label{eriocc}
\end{align}
where MO transformed multipole integrals are defined as
\begin{subequations}
\begin{align}
&O_{l,m}^{rj,\mathbf{X}} = \sum_{s} O_{l,m}^{rs,\mathbf{X}} C_{sj},\\
&O_{l,m}^{ij,\mathbf{X}} = \sum_{r} O_{l,m}^{rj,\mathbf{X}} C_{ri}.
\label{full}
\end{align}
\label{motrans}
\end{subequations}
It is important to stress that one of the multipoles in Eq.~\eqref{eriocc} is fully transformed to the MO basis; therefore, its size remains the same
at the coarse-grained level, allowing us to evaluate it using the FMM algorithm with a hierarchy of boxes.

The traditional FMM algorithm is modified as follows (see graphical explanation in Fig.~\ref{algfig}).
First, for each box at the lowest level, we compute $O_{l,m}^{rs,\mathbf{X}}$ and transform them to the MO basis, $O_{l,m}^{ij,\mathbf{X}}$, using
Eq.~\eqref{motrans}.
We then compute $\mathscr{M}^{ij}_{l,m}(\mathbf{X})$ that is analogous to Eq.~\eqref{fmmc},
\begin{align}
\mathscr{M}^{ij}_{l,m}(\mathbf{X}) = (-1)^l\sum_{\mathbf{X}'} \sum_{l'm'}M_{l+l',m+m'}(\mathbf{X}-\mathbf{X}') O^{ij,\mathbf{X}'}_{l',m'}.
\label{fmmex}
\end{align}
Note that the summation over $\mathbf{X}'$ in this equation is essential for utilizing the translations in FMM discussed above.
When computing $\mathscr{M}^{ij}_{l,m}(\mathbf{X})$, we use the same algorithm as the traditional FMM, namely those based on Eqs.~\eqref{fmm1}, \eqref{fmm2}, and \eqref{fmm3}
for each pair of $i$ and $j$:
\begin{subequations}
\begin{align}
&O_{l,m}^{ij,\mathbf{X}_p} = \sum_{l'm'} O_{l-l',m-m'}^{ij,\mathbf{X}_c - \mathbf{X}_p} O_{l',m'}^{ij,\mathbf{X}_c},\\
&\mathscr{M}_{l,m}^{ij}(\mathbf{X}) = (-1)^l \sum_{l'm'}M_{l+l',m+m'}(\mathbf{X} - \mathbf{X}') O_{l',m'}^{ij,\mathbf{X}'},\\
&\mathscr{M}_{l,m}^{ij}(\mathbf{X}_c) = \sum_{l'm'}M_{l',m'}(\mathbf{X}_p) O_{l'-l,m'-m}^{ij,\mathbf{X}_c - \mathbf{X}_p}.
\end{align}
\end{subequations}
The occupied-orbital exchange matrix is then computed as
\begin{align}
K_{ri}^\mathrm{ff} = \sum_{j} \sum_{lm} O_{l,m}^{rj,\mathbf{X}} \mathscr{M}_{l,m}^{ij}(\mathbf{X}).
\label{kri}
\end{align}
The near-field contributions to the occupied-orbital exchange matrix
can be computed simultaneously with those to the Coulomb matrix with marginal additional costs.

There are a number of parameters required to perform FMM calculations, and some are dependent on the system of interest. 
The number of levels or depth ($N_s$) in FMM is typically chosen to be 4 or 5, such that the length of the smallest box
is about 2.0 bohr.
This number determines the total number of boxes as well as the size and number of boxes
at the lowest level, and therefore, affects the efficiency of FMM.
The definition of the near- and far-field regions depends of a number of parameters, for which interested readers can refer to
Refs.~\onlinecite{white94cpl}, \onlinecite{choi01jcc}, and \onlinecite{strain96science}.
Our implementation makes explicit use of contracted basis functions
to optimize the computation of the electron repulsion integrals in the near-field region.
The definition of the extent of each distribution used to determine the near- and far-field regions is
described in Refs. \onlinecite{perez-jorda97jcp} and \onlinecite{sierka03jcp}. 
The `well-separatedness' index ($ws$) is typically chosen to be 0 such that two charge distributions are
considered non-overlapping if the distance between their centers is greater than the sum of their extents.
However, this parameter can be tuned depending on the definition of the extents and the systems studied.

\section{Numerical Results}
In this section, we first show the convergence of the Coulomb and exchange energy contributions with respect to multipole ranks.
We then present the parallel scaling of our algorithm, followed by the timing data
using the optimized parameters. 

\subsection{Convergence with respect to multipole ranks}
\begin{table*}[tb]
\caption{Convergence of the energy with respect to multipole ranks $L_\mathrm{max}^J$ and $L_\mathrm{max}^K$ for the 
         graphene sheet C$_{96}$H$_{24}$ using the def2-SVP basis set.
         Errors are shown in $\mathrm{m}E_\mathrm{h}$ with respect to the reference energy
         computed using $L_\mathrm{max}^J=15$ and $L_\mathrm{max}^K=5$ ($-3647.27031641$~$E_\mathrm{h}$).}
\label{convergence}
\begin{ruledtabular}
\begin{tabular}{crrrrrr}
  $L_\mathrm{max}^J$ & {$\mathrm{no\ exchange}$} & {$L_\mathrm{max}^K=0$} & {$L_\mathrm{max}^K=1$} & {$L_\mathrm{max}^K=2$}
                                            & {$L_\mathrm{max}^K=3$} & {$L_\mathrm{max}^K=4$} \\
  \hline
  0  &  116841.178  &   116844.491 &  116844.849 & 116844.854 &  116844.853  &  116844.853 \\
  1  &  5032.828    &  5036.141    &  5036.499   & 5036.504   &  5036.504    &  5036.504 \\
  2  &  285.992     &  289.304     &  289.662    & 289.668    &  289.667     &  289.667 \\
  3  &  25.565      &  28.877      &  29.235     & 29.241     &  29.240      &  29.240 \\
  4  &  $-$0.063    &  3.249       &  3.607      & 3.613      &  3.612       &  3.612 \\
  5  &  $-$3.094    &  0.218       &  0.576      & 0.582      &  0.581       &  0.581 \\
  6  &  $-$3.604    &  $-$0.291    &  0.067      & 0.072      &  0.072       &  0.072 \\
  7  &  $-$3.669    &  $-$0.357    &  0.001      & 0.007      &  0.006       &  0.006 \\
  8  &  $-$3.675    &  $-$0.362    &  $-$0.004   & 0.001      &  0.001       &  0.001 \\
  9  &  $-$3.675    &  $-$0.363    &  $-$0.005   & 0.001      &  0.000       &  0.000 \\
  10 &  $-$3.675    &  $-$0.363    &  $-$0.005   & 0.001      &  0.000       &  0.000 \\
\end{tabular}
\end{ruledtabular}
\end{table*}

We examined the convergence of the Hartree--Fock energy with respect to the ranks of multipole expansions, $L_\mathrm{max}^J$ and $L_\mathrm{max}^K$,
for a graphene sheet C$_{96}$H$_{24}$.
We chose this system because the exchange contributions in graphene sheets have been shown to be slowly decaying with distance.\cite{burant96jcp}
The def2-SVP basis set was used. 
We set the FMM parameters to be $N_s=5$, $ws=-0.1$. The Schwarz integral screening and 
SCF convergence thresholds were set to $1.0\times 10^{-8}$.
In the reference calculation, the multipole expansions
were truncated at $L_\mathrm{max}^J = 15$ and $L_\mathrm{max}^K=5$ for the far-field Coulomb and exchange interactions, respectively.
The convergence was analyzed by comparing the reference energy
and that computed from the Fock operator constructed using
the reference MO coefficients and
different values for $L_\mathrm{max}^J$ and $L_\mathrm{max}^K$. The results are shown in Table \ref{convergence}.
The errors decay quickly for both the Coulomb and exchange contributions as higher-rank multipoles are included.
However, since the magnitude of the far-field exchange contribution ($\sim4\ mE_\mathrm{h}$) is a few orders of magnitude smaller than
that of the far-field Coulomb contribution,
$L_\mathrm{max}^K$ can be smaller than $L_\mathrm{max}^J$, 
thus significantly reducing the computational cost at almost no loss in accuracy.
It is worth noting that the error in the far-field exchange contributions is around $1\ \mu E_\mathrm{h}$ with $L_\mathrm{max}^K=2$ for this challenging system.
From these results, we concluded that the multipole series should be truncated at $L_\mathrm{max}^J=10$ 
for the Coulomb interaction and at $L_\mathrm{max}^K=2$ for the exchange interaction to achieve $\mu E_\mathrm{h}$ accuracy.

\subsection{Parallel scalability}
\begin{figure}[t]
\includegraphics[width=0.5\textwidth]{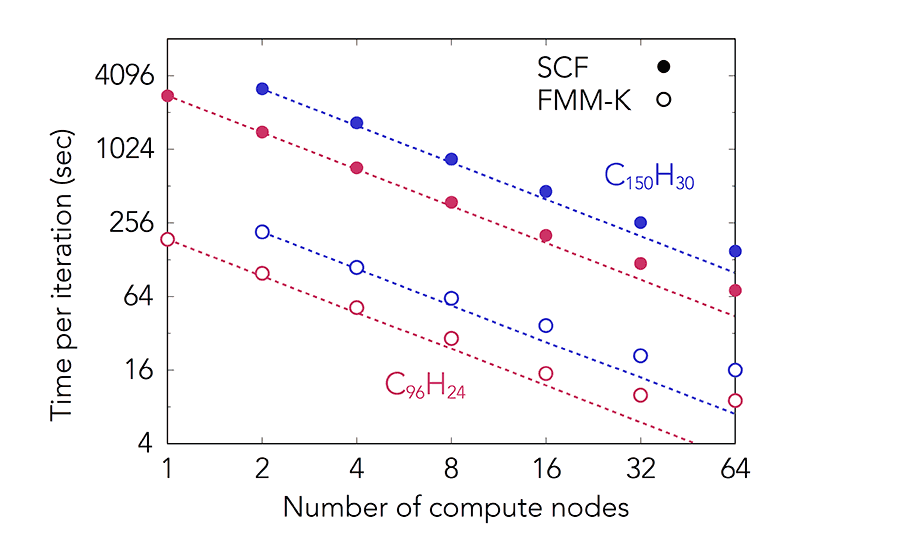}
\caption{Timings for graphene sheets C$_{96}$H$_{24}$ and C$_{150}$H$_{30}$ using def2-SVP and the parameters NS $=5$, $L_\mathrm{max}^J=10$,
         and $L_\mathrm{max}^K=2$. Each compute node consists of 2 Xeon E5-2650 CPUs (Sandy Bridge 2.0GHz).}
\label{scaling}
\end{figure}

Our algorithm can be trivially parallelized with very high efficiency, making it useful for large-scale
problems.
We measured the strong parallel scaling using the graphene sheets C$_{96}$H$_{24}$ and C$_{150}$H$_{30}$. 
The FMM parameters used for these calculations were $N_s=5$, $ws=0$, $L_\mathrm{max}^J=10$, and $L_\mathrm{max}^K=2$.
The Schwarz integral screening threshold was set to $1.0\times 10^{-8}$.
The results are shown in Fig.~\ref{scaling}. Calculations were performed using the def2-SVP basis set on a 64-node computer cluster,
where each compute node consists of 2 Xeon E5-2650 CPUs (Sandy Bridge 2.0GHz). Total timings for an
SCF iteration and timings for the far-field exchange evaluation were averaged over the first 5 iterations.
The cost of the far-field exchange evaluation for C$_{96}$H$_{24}$, which was about 10\% of the total cost per SCF iteration,
was 188 sec with 1 compute node, and reduced to 99, 52, 29, 15, 10, and 9 sec
using 2, 4, 8, 16, 32, and 64 compute nodes.
The timings for C$_{150}$H$_{30}$ were 216, 111, 62, 37, 21, and 16 sec using 2, 4, 8, 16, 32, and 64 compute nodes.
For C$_{96}$H$_{24}$, the strong scalings from 1 to 64 compute nodes for an SCF iteration and far-field exchange evaluation were found to be 61\% and 33\%, respectively.
Those for C$_{150}$H$_{30}$ from 2 to 64 nodes were 66\% and 42\%.

The excellent scaling for the far-field exchange evaluation is due to the fact that the transformation of the multipole tensors from the AO basis $O_{l,m}^{rs, \mathbf{X}}$
to the occupied MO basis $O_{l,m}^{ij, \mathbf{X}}$ in Step 2 of the occ-FMM-K algorithm (Fig.~\ref{algfig}) 
can be done independently in batches of occupied-orbital index $j$.
As a result, the upward and downward translations of the multipoles and local expansions in the occupied MO basis in Step 3--5 are well distributed.
The construction of the partial exchange matrix from the multipoles and local expansions (Eq.~\ref{kri}) is also similarly parallelized.
The near-field Coulomb and exchange contributions are calculated with exact four-center integrals at the moment and
accounts for most of the differences between the total timing for an SCF iteration and the time taken for the far-field exchange evaluation.

\subsection{Timing data}

\begin{figure}[tb]
\includegraphics[width=0.5\textwidth]{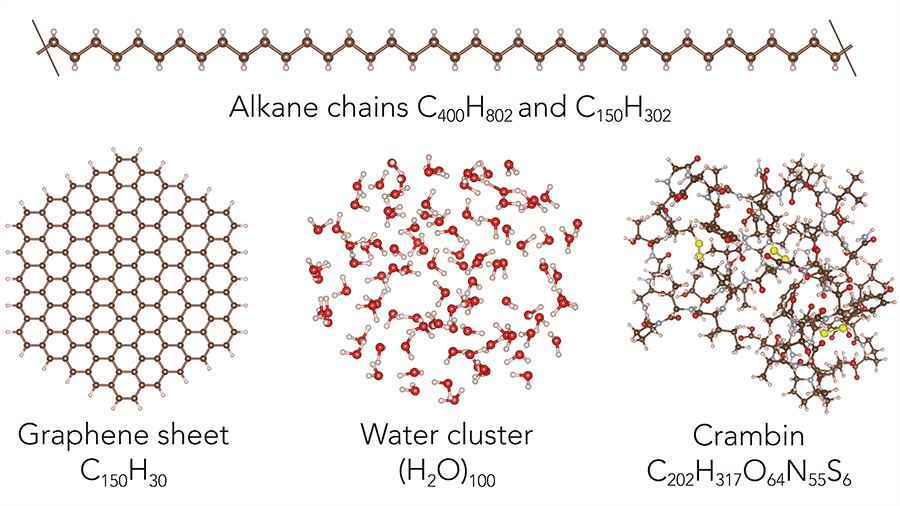}
\caption{Systems used for timing benchmarks in this work.}
\label{fgr:system}
\end{figure}

\begin{table*}[tb]
\caption{ Wall time (min) for calculating the far-field exchange and Coulomb contributions using the FMM algorithms.
The total timing for an SCF iteration is also shown.
128 Xeon E5-2650 CPUs (Sandy Bridge 2.0GHz, total 1024 cores) with InfiniBand QDR were used.}
\label{timingtab}
\begin{ruledtabular}
\begin{tabular}{llrrrrrr}
  System     &   & Atoms & Electrons & Basis\footnotemark[1] & Far-field K & Far-field J & SCF iter. \\
  \hline
  Alkane chain   & C$_{150}$H$_{302}$                        & 452      & 1202       & 3610       & 0.1        & 0.0    & 0.6  \\
  Alkane chain   & C$_{400}$H$_{802}$                        & 1202     & 3202       & 9610       & 0.3        & 0.1    & 3.5  \\
  Graphene sheet & C$_{150}$H$_{30}$                         & 180      & 930        & 2250       & 0.3        & 0.1    & 2.5  \\
  Water cluster  & (H$_2$O)$_{100}$                          & 300      & 1000       & 2400       & 0.8        & 0.5    & 2.5  \\
  Crambin        & C$_{202}$H$_{317}$O$_{64}$N$_{55}$S$_6$   & 644      & 2522       & 6187       & 1.4        & 0.2    & 41.2 \\
\end{tabular}
\end{ruledtabular}
\footnotetext[1]{The def2-SVP basis set was used.}
\end{table*}

The performance of our occ-FMM-K implementation is assessed for a number of molecular systems (shown in Fig.~\ref{fgr:system}) using
the def2-SVP basis set.
The results are compiled in Table~\ref{timingtab}.
The parameters used in all of the timing benchmark calculations 
are $N_s = 5$ (except for the water cluster for which we used $N_s=4$), $ws = 0$, $L_\mathrm{max}^J=10$, and $L_\mathrm{max}^K=2$.
The Schwarz integral screening threshold was set to $1.0\times 10^{-8}$.
In principle, the sets of the FMM parameters used for the evaluation of far-field Coulomb and exchange contributions
can be different. We have not yet investigated how the parameters
besides $L_\mathrm{max}$ can be optimized to achieve maximum efficiency without loss of accuracy.
It is, however, expected that the optimal parameters used for the Coulomb interaction  will be 
different from those used for the exchange interaction
as the Coulomb interaction is longer-range, and the cost of evaluating
the Coulomb contribution is significantly smaller.
This will be investigated in the future.

We included the one-dimensional alkane chains C$_{150}$H$_{302}$ and C$_{400}$H$_{802}$ as examples, because
FMM is known to perform most efficiently for one-dimensional systems (even though the far-field exchange contributions to the total energies for these particular systems are negligible).
This efficiency is due to the fact that the fraction of boxes in the near field does not change with system size in one-dimension.
From a 100-carbon chain (3610 basis functions) to a 400-carbon chain (9610 basis functions), the total timing for
an SCF iteration increased from 0.6 min to 3.5 min.
In both cases, the cost of computing the far-field exchange contribution was only a fraction of that for the near-field contributions;
the far-field exchange computation amounted to 17\% (in C$_{150}$H$_{302}$) and 9\% (in C$_{400}$H$_{802}$)
of the total timing per SCF iteration, respectively.
The cost of computing the far-field Coulomb contribution was also small. 

Next we performed a calculation for a two-dimensional graphene sheet C$_{150}$H$_{30}$.
As mentioned previously, this is considered among the most challenging systems for exchange computation, because the exchange interaction decays slowly with distance.
Note that this example was the largest two-dimensional system used in the benchmarks by Burant and Scuseria\cite{burant96jcp}
for their NFX method that accounts for the far-field exchange contributions by simply increasing the size of the near-field FMM. For this example, the
cost of the far-field exchange evaluation using our algorithm was around 0.3 min, which was 12\% of the total cost for an SCF iteration (2.5 min).
The remaining cost is largely due to the near-field four-center integral evaluation and diagonalization of the Fock matrix.

Finally, the timings are reported for a water cluster (H$_2$O)$_{100}$ (Ref.~\onlinecite{supp})
and a small protein Crambin C$_{202}$H$_{317}$O$_{64}$N$_{55}$S$_6$ to assess the performance of our algorithm for three-dimensional systems.
The latter was previously used to benchmark the DFT and DLPNO-CCSD(T) methods.\cite{riplinger13jcp, riplinger16jcp, furche06jcp}
The cost of far-field exchange evaluation was 32\% and 3\% of that of an SCF iteration for the water cluster and Crambin, respectively.
Similar to the previous examples, a large portion of the the remaining cost is attributed to the near-field four-center integral evaluation.

These results, together with the excellent parallel scaling of our algorithm, are highly encouraging. It is also worth noting that (1) the cost of the near-field computation
can be further reduced using, for example, the RI approximation;
and (2) the use of localized orbitals and screening of occupied-orbital pairs would significantly reduce the cost of the far-field exchange evaluation.
The memory requirement for large calculations is determined at the moment by the size of the multipole and local expansion tensors
for each box at the lowest level.

\section{Conclusion}
In this paper, we introduced an efficient FMM-based algorithm for evaluating the exact exchange matrix elements that contribute to the 
energy and orbital-rotation gradient at the mean-field level. This is done by constructing the partial exchange matrix $K_{ri}$,
where all matrix elements have at least one occupied-orbital index. The multipole and local expansion tensors are first transformed
into the occupied-orbital basis. The upward and downward translations of these tensors are then performed in exactly the same
manner as conventional FMM for the Coulomb interaction. Efficient parallelization
and the fact that there is no assumption on the sparsity of the density matrix make this algorithm attractive
for large and extended systems, especially those with small band gaps.

There are, however, a number of ways to further improve our algorithm.
First, it is possible to reduce the cost of the far-field exchange evaluation for many systems by using localized molecular orbitals and screening occupied-orbital pairs.
This would significantly mitigate the cost of storage and basis transformations.
Second, the expensive near-field integral evaluation can be replaced by an algorithm based on the RI approximation.
In addition, extensions of our algorithm to complete active space self-consistent field (CASSCF) and configuration interaction singles (CIS) should be straightforward.
These improvements and extensions will be investigated in the near future.

\begin{acknowledgments}
We thank Dr. Jae Woo Park for providing the geometry of the water cluster.
This work has been supported by National Science Foundation ACI-1550481 (HAL) and CHE-1351598 (TS). 
T.S. is an Alfred P. Sloan Fellow.
\end{acknowledgments}

\end{document}